# A Transparency Index Framework for AI in Education

*Muhammad Ali Chaudhry*
UCL Knowledge Lab, muhammad.chaudhry.16@ucl.ac.uk

*Mutlu Cukurova*
UCL Knowledge Lab, m.cukurova@ucl.ac.uk

*Rose Luckin*
UCL Knowledge Lab, r.luckin@ucl.ac.uk

Numerous AI ethics checklists and frameworks have been proposed focusing on different dimensions of ethical AI such as fairness, explainability, and safety. Yet, no such work has been done on developing transparent AI systems for real-world educational scenarios. This paper presents a Transparency Index framework that has been iteratively co-designed with different stakeholders of AI in education, including educators, ed-tech experts, and AI practitioners. We map the requirements of transparency for different categories of stakeholders of AI in education and demonstrate that transparency considerations are embedded in the entire AI development process from the data collection stage until the AI system is deployed in the real world and iteratively improved. We also demonstrate how transparency enables the implementation of other ethical AI dimensions in Education like interpretability, accountability, and safety. In conclusion, we discuss the directions for future research in this newly emerging field. The main contribution of this study is that it highlights the importance of transparency in developing AI-powered educational technologies and proposes an index framework for its conceptualization for AI in education.

## 1. INTRODUCTION

Ethical AI is a rapidly developing field with a number of tools, frameworks and research papers coming out at a high frequency. Considering the blurred boundaries between different dimensions of ethical AI like fairness, accountability, transparency and explainability, it is important to first define what we mean by transparency in the context of AI. Both, AI and transparency can be interpreted in many ways (Weller, 2017; Felzmann, 2019). In the context of this study, transparency in AI refers to a process with which all the information, decisions, decision-making processes and assumptions are made available to be shared with the stakeholders and this shared information enhances the understanding of these stakeholders.

Transparency, according to Turilli and Florodi (2009), depends on factors such as how readily accessible information is, and how it can be used pragmatically to aid decision-making. The second part of this definition which emphasizes enhancing the understanding of stakeholders shows that transparency is inherently dependent on the people for whom it is targeted. It also means that a construct that might be considered transparent for one person might be a black box for another. In AI for education, a transparent product development pipeline for an AI practitioner might be a complete black box for an end-user like an educator who is not a tech expert but is impacted by that product.

AI can empower the human agency with more informed decision-making. Floridi (2018) argues that the agency humans enjoy and the extent to which they delegate is not zero-sum. AI has the potential to multiply human agency. But, for AI to empower this human agency and to ensure smooth human-AI interaction, transparency is essential.

While there has been a lot of work on ethical AI (i.e Kazim and Koshiyama, 2020; Jobin et al, 2019), the focus on transparency as a necessary construct to enable ethical AI is limited. More specifically in education, it is almost non-existent. The research presented in this paper aims to fill in this gap by presenting and evaluating a framework that tackles the challenges of ethical AI through transparency in the entire product lifecycle of AI-powered ed-tech products, from initial planning of an AI system till it is deployed and iteratively improved in the real world.

The goal of this research is to explore what kind of design framework can be applied to ensure transparency in the development of AI-powered educational technology. Taking transparency as a subjective construct that is dependent on the person for whom it is targeted, we also explore how the requirements of transparency evolve with different stakeholder groups in educational contexts. Lastly, we investigate the extent to which transparency in the AI development process enables the other dimensions of ethical AI like explainability and interpretability, fairness, accountability and safety.

We describe our research as a co-design methodology of a framework on the transparency of AI in education. Our research began with a review of literature on the different stages of the standard AI development pipeline and identified popular domain-agnostic frameworks for describing each stage: data processing, machine learning modelling, deployment, and iterative improvements. These frameworks were then applied in a real-world setting while developing AI powered educational tools for a training organization. During this process, all the requirements for transparency of AI in educational contexts were coherently merged into a single framework that is presented in this paper as the Transparency Index. The framework was then evaluated with different stakeholders of education (educators, ed-tech experts and AI practitioners) in two phases and improved iteratively.

The Transparency Index framework proposed in this research focuses on transparency from a pragmatic viewpoint in the planning, development and deployment of AI tools that are used in learning contexts. This product development lifecycle starts from initial discussions on the purpose of an AI tool, its deployment strategy and the decision to start collecting the data. It does not end after deployment but rather is a continuous process. The framework created in this study covers all these aspects as data transparency, algorithmic transparency and deployment transparency of AI products that are used in learning contexts.

### 1.1 Transparency and other aspects of ethical AI in education

One of the key strengths of ensuring transparency in the AI development process is that it has a significant overlap with other ethical AI dimensions. There are many ways in which transparency informs and enables other dimensions of ethical AI to be implemented during the AI development process. For example, one of the requirements of the Algorithmic Transparency stage in the Transparency Index framework is for an ed-tech company to share the tools that they are using to implement explanations for their AI system. This would ensure explainability. User testing in the 'Implementation Transparency' section of the Transparency Index framework would ensure that relevant stakeholders of an AI-powered ed-tech product understand these explanations and their limitations. Hence, enabling interpretability.

Figure 1 below shows how transparency as conceptualized in the Transparency Index framework can overlap with explainability, fairness, accountability, interpretability and safety of AI powered ed-tech systems.

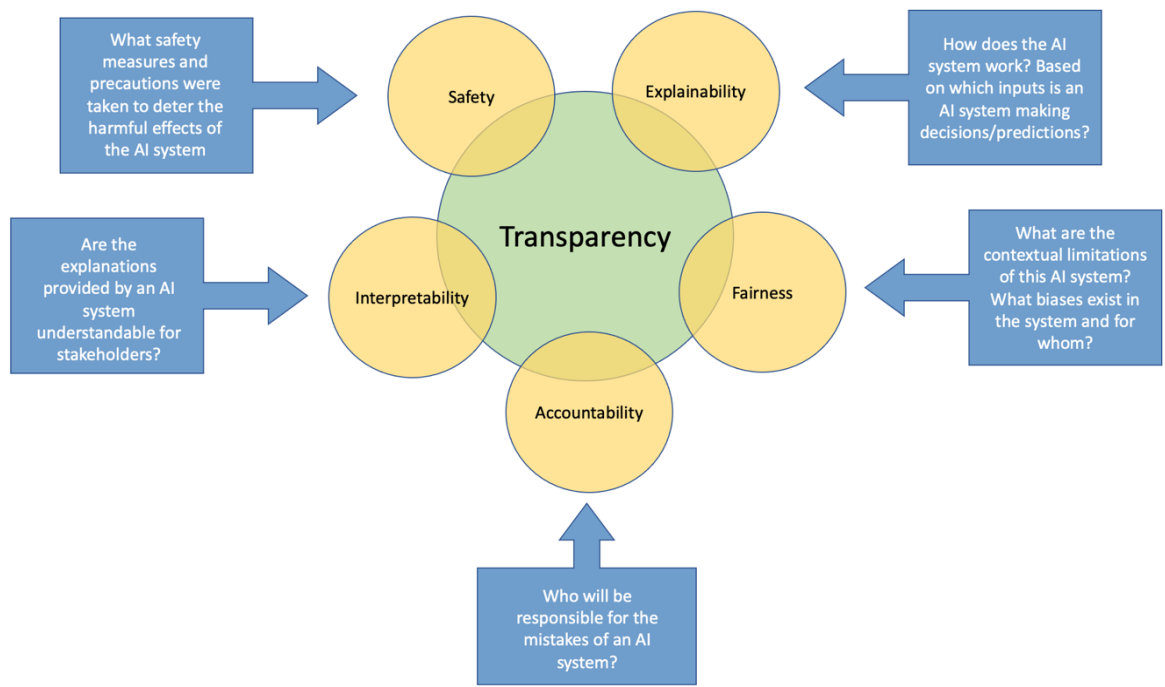

*Figure 1: Transparency in relation to the other dimensions of ethical AI*

Explainability and Interpretability of AI systems are at times used interchangeably to illustrate how easy it is for humans to understand the cause of a particular decision by an AI system (Linardatos et al, 2021). They seem to be necessary prerequisites for transparent AI systems as well. Ed-tech companies do not need to share all the details of their AI tools with end-users. But the information they do share needs to be understandable by the users of the AI system as highlighted in the Transparency Index framework.

According to some researchers, transparency and accountability are closely related (Hood, 2010; Matthias, 2004). From the viewpoint of ed-tech companies, there are strong reasons why accountability leads to more transparent AI development processes and vice versa. Accountability in terms of transparency for ed-tech companies is a two-edged sword. On one hand, ed-tech companies might want to make their development processes transparent and share all the details (pros and cons) with educational institutions to avoid taking any responsibility for mishaps. But, on the other hand they may want to avoid sharing any details with educational institutions because it makes their product difficult to sell and they might be held accountable for the weaknesses or misuse of their products. Transparency as depicted in the Transparency Index framework for AI-powered ed-tech can help companies document and test the assumptions on which they build their AI systems and hold the relevant departments or individuals accountable when tools perform unexpectedly.

Within different dimensions of ethical AI, safety is of utmost importance and cannot be compromised. It quite often gets ignored in the race to be the first one to launch an AI-powered product. In education, safety is even more crucial because of high stakes and because mishaps in AI-powered ed-tech can go unnoticed unless a teacher or student raises it with the relevant authorities in their school. The safety of AI systems, like transparency, is not dependent on a particular tool but needs to be ingrained in the whole development process. The transparency Index framework presented in this research can lead to more robust AI systems because it encourages the ed-tech providers to document the details of their data processing, machine learning modelling, user testing and deployment stages. This documentation enables AI practitioners to test their assumptions, justify their decisions and adopt a more cautious approach when selecting different technical tools, libraries and third-party services in building their products. Hence, leading to more safe AI systems.

Theoretically, transparency in AI does not replace ethical AI, it is its subset. But if there is one dimension of ethical AI that AI practitioners need to choose to focus on, it can potentially be transparency in AI for a number of reasons. Firstly, transparency covers the entire AI development process from initial designs till deployment in the real-world as it co-exists with the AI tool. Secondly, it can ensure (as seen from the Transparency Index framework later) that the other ethical AI dimensions like explainability, safety and fairness are being addressed as well. Thirdly, it facilitates and benefits the two major stakeholders of an AI-powered ed-tech: firstly ed-tech companies through a thorough documentation of the tool and robust development processes, and secondly, end-users through a better understanding of how the AI system works.

## 2. LITERATURE REVIEW

In this section we review the literature on the role of transparency in AI in general and then cover the latest research on transparency in AI for Education.

Transparency in AI is not a new construct (Larsson and Heintz, 2020). It is at the centre of ethical AI (Woudstra, 2020). It is not an instantaneous phenomenon that is dependent on a particular decision, but a continuous process that accompanies the entire AI product development lifecycle.

Developing an AI tool is a complex, time consuming and resource-intensive process. The very first decision to build an AI tool and define its operations involves assumptions that can be challenged or changed. Irrespective of the sector in which AI is applied, transparency is essential to enhance the understanding of relevant stakeholders regarding questions like how the AI works, what are its limitations, in which contexts should it be avoided and how does it improve *the status quo*.

De Fine *et al.* (2020) have simplified the concept of transparency in AI tools by dividing the development process of AI tools into three phases:
1. Goal setting to define the tasks undertaken by the AI
2. Coding to take account of the development process of an AI tool
3. Implementation to illustrate how the tool is used in real-world applications

It is intuitive that transparent AI tools might be preferred over less transparent ones, but there is no guarantee that users will always prefer complete transparency in AI tools. No matter how transparent a particular AI implementation is or how much information is shared with the end-users, there is no easy way for end-users to find out if a company has hidden any information. This means there is always a possibility that end-users may not trust the system, irrespective of any transparency or ethical measures taken by the company.

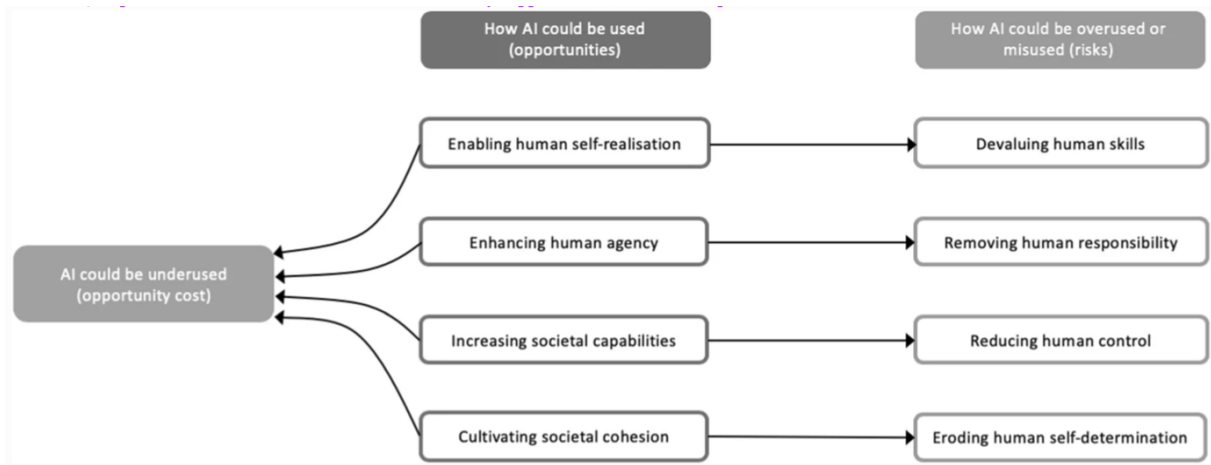

*Figure 2: Overview of the opportunities and risks offered by AI and their opportunity cost (Floridi et al, 2018)*

Figure 2 above from Floridi et al (2018) shows the opportunities and risks offered by AI applications. Transparency is at the heart of any measures we take to mitigate these risks. It is the first step towards enhancing the understanding of AI by humans who can potentially face devalued skills, removed responsibility, reduced control and eroded self-determination due to AI, as shown in the Figure 2 above.

### 2.1 Transparency of AI in General

AI is hugely impacting the way we learn (Luckin, 2018), stay healthy (Hansel et al, 2015), cure diseases (Shen et al, 2019), spend money (Smith and Linden, 2017), maintain order in our societies (Brayne and Christin, 2020) and take organizational decisions (Jarrahi, 2018; Philips-Wren, 2012; AlgorithmWatch, 2019). This penetration of AI in our daily lives has also magnified the risks it poses (Andrew et al, 2019).

Recently, there has been significant research and adoption of ethical AI principles like fairness, accountability, interpretability and explainability (European Commission, 2020; IEEE, 2019). This is driven by firstly the impact of AI applications in our daily lives (Bughin et al, 2018; Crawford et al, 2016; Vaishya, 2020), and secondly, the mishaps of AI systems in the real world (Kaushal et al, 2020; Wellner, 2020; Raji and Buolamwini, 2019; Zavrsnik, 2020). There have been a number of tools, checklists and frameworks published to take account of ethical considerations in developing, deploying and auditing AI products (Morley, 2020; Dameski, 2018; Leikas, 2019; Winfield, 2019; Deepmind Safety Research, 2018) but there is not much work focusing on transparency in particular throughout the AI's planning, development and deployment stages.

This lack of research is problematic. An AI tool built with huge amounts of data and the best performing machine learning algorithms will perform at its best only in certain contexts. Transparency is essential to know in which contexts the tool will not perform at its optimal level. It is widely accepted that bias or discrimination cannot be completely removed from an AI tool (IBM, 2018; ICDPPD, 2018), but we can mitigate some of them. The extent to which bias or discrimination exists in a particular AI product, applied in a certain context with a particular type of users can only be determined if the details of the tool's development are documented and shared in a transparent manner.

For instance, Felzmann *et al.* (2020) have proposed a framework for transparency through the design of AI systems. The purpose of their framework is to bridge the gap between high-level AI ethics principles and AI practitioners who are developing AI tools. One of the principles in their framework focuses on transparency as an integrative process throughout the AI tool's development pipeline.

Richard and King (2013) have identified transparency among the three paradoxes of big data in AI. AI aims to make the world more transparent and AI tools in different sectors like healthcare, education and governance claim to empower the individuals, but they seem to be doing this in a very secretive manner where decisions and assumptions are not documented (Holstein et al, 2019), machine learning models used are opaque (Castelvecchi, 2016) and the limitations of these tools are not shared (Besold, 2014).

Ananny and Crawford (2016) have discussed the limitations of transparency in ensuring or guaranteeing ethical AI. They evaluate transparency at two levels: algorithmic transparency in the code that AI practitioners write (Pasquale, 2015; Diakopoulos, 2016; Brill 2015; Dubber et al, 2020) and design transparency regarding how the AI systems are planned, developed, deployed and evaluated (Hollanek, 2020, Plale, 2019; Wischmeyer, 2020). Some of the issues raised by these researchers are discussed in the discussion section later.

Education is a high-impact domain where a wrong prediction by an AI system can confuse teachers and have adverse psychological impact on learners. Hence, the lack of research on the Transparency of AI in educational products is concerning and needs to be addressed.

### 2.2 Transparency of AI in Education

Transparency of AI products in education is a relatively new concept compared to other sectors like healthcare, recruitment or justice system where AI-powered products are being widely used. Within AIED, the transparency of AI-powered products needs to be considered within the context of ethics of education like what is the purpose of learning (to pass exams or help students achieve self-actualization), what is the role of AI (to empower teachers or replace them) and will AI be creating equal access to education or widen the gap between affluent and disadvantaged communities (Holmes, 2019a and 2019b). There has been a lot of work to align the goals of education and technology (Luckin, 2021; Cukurova et al, 2019; Moeini, 2020)

Recently, there has been a growing interest of AI researchers and practitioners within education on the ethical implications of AI products on learners and teachers. Holmes et al (2021) have presented a community-wide framework for developing ethical AI for education. Figure 3 below shows three major components of their framework for developing AI in education: education, algorithms and big data.

For ethical AI, there needs to be a free flow of information between experts involved in bringing the three distinct components shown in figure 3 together. Transparency can play a significant role in achieving this. For educators, it can ensure that the purpose of an AI tool is clearly specified and conveyed to ed-tech experts who then convey it to AI practitioners. For ed-tech experts, transparency ensures they clearly understand the requirements of educators, the ed-tech being used in their schools and the quality of data that can be collected from their ed-tech. For AI practitioners, transparency can ensure that they clearly understand the needs of educators, strengths and weaknesses of the data in hand and documentation of the accuracy metrics of machine learning models.

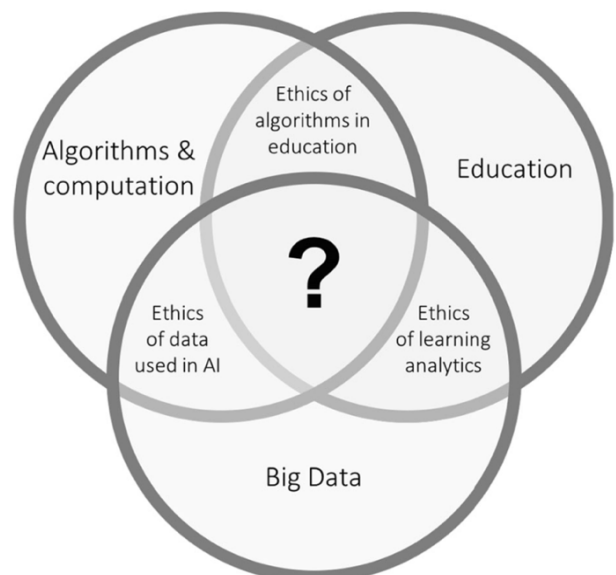

*Figure 3: A draft framework for the ethics of AIED*

The Institute for Ethical AI in Education (2021) collaborated with over 200 experts in AIED through interviews, roundtables and The Global Summit on the Ethics of AI in Education to prepare an ethical framework for AI in education that provides detailed criteria and a checklist to educators for evaluating AI-powered ed-tech products. Their framework aims to empower the 'leaders and practitioners to drive the design, procurement and application of AI on behalf of learners'. They also identify 'transparency and accountability' as one of the objectives that educators need to look for when selecting AI products for their educational institutions.

Transparency in AI systems is a continuous process that goes on throughout the AI development pipeline. Considering the complexities of the machine learning development pipeline, researchers have produced a number of frameworks to document its different stages. For example, Gebru et al (2018) introduced 'datasheets for dataset' to standardize the documentation of datasets (including their strengths and weaknesses) used to train machine learning algorithms. Mitchell et al (2019) have introduced model cards to document the strengths and weaknesses of machine learning models used to make predictions. Arnold et al (2019), from IBM have produced FactSheets to facilitate AI service providers in documenting their products' functionalities, performance, safety and security. Although relevant attempts to bring transparency in AI has been made in various contexts, currently there is no framework to pull them together for their relevance in the context of educational applications

Ethical AIED is an emerging field with very limited peer-reviewed publications in the literature. The work on transparency as a pivotal dimension of ethical AI for education is almost non-existent. This paper aims to bridge this gap. The Transparency Index framework proposed in this research brings available contributions to ethical AI in general together under a single framework in the context of education to ensure transparency throughout the design, development and deployment of AI-powered ed-tech products. The contributions of this paper are threefold: Firstly, it presents a Transparency Index framework to support the ethical development and implementation of AI in Education. Secondly, it evaluates the presented framework through interviews with different stakeholder groups within education: educators, ed-tech experts and AI practitioners. This evaluation shows if, and how this framework can benefit these stakeholders. Thirdly, the interviews with educators, ed-tech experts and AI practitioners for evaluating the framework reveal some interesting dynamics between these three key stakeholders that can have significant implications for ethical development and deployment of AI in Education.

## 3. METHODOLOGY

A mixed methods approach was used to first develop, then evaluate and improve the framework developed in this study. First, the framework was developed based on a thorough literature review of the standard machine learning development pipelines for AI systems and an application of the shortlisted frameworks on different aspects of the AI development process for educational contexts. The second phase involved evaluating and iteratively improving the framework based on qualitative data collected from the interviews of education stakeholders.

### 3.1 Framework Creation

The framework was created in two steps: firstly, based on the literature review of different stages of the AI development process: data processing stage, machine learning modelling stage and deployment and iterative improvements stage. In this step, popular frameworks for the documentation, robustness and reproducibility for each of these stages were identified.

In the second step, the selected framework for each stage of the AI development process was applied in the real world during the development of an AI tool in an educational context for a training organization that envisioned to become a leader in educational technology. This application of domain agnostic frameworks for AI tool development in an educational context enabled us to identify any gaps in these frameworks, when applied in educational settings.

For the data processing stage, datasheets from Gebru et al (2018) were chosen as the benchmark for documenting different components of the data processing stage in the AI development process. In the Transparency Index framework, datasheets were wrapped around other requirements for the data processing stage to make it more applicable and suitable for educational contexts.

For the Machine Learning modelling stage, model cards by Mitchell et al (2019) were chosen as a baseline to document the details of the ML model used. Some additional requirements were also added for this stage to record the various decisions and assumptions made specifically for the AI-powered ed-tech products. These requirements were derived from our experience of applying AI to enhance the understanding of domain experts (Kent et al, 2021).

Factsheets by Arnold et al (2019) were added as a basic requirement for documenting the usability of AI tools. Considering the importance of user feedback in ensuring the effectiveness of AI tools in educational contexts, some additional requirements were also merged with factsheets for the deployment and testing of AI-powered ed-tech products in real-world settings.

AI development is a complex and time-consuming process. All these frameworks cover different stages of the AI development process. They were brought together in a coherent manner under the umbrella of the Transparency Index framework and were accompanied by some other requirements to specifically suit the needs of different stakeholders in education.

### 3.2 Framework Evaluation

#### 3.2.1 Participants

The framework was iteratively evaluated in two phases with three groups of educational stakeholders. 40 candidates were recruited in two phases to participate in this research. 18 candidates responded to this request and participated in interviews. These candidates were divided into three groups and short-listed based on their backgrounds, as shown in Table 1 below. Ten candidates were interviewed in phase 1 and eight candidates were interviewed in phase 2. Nine participants requested a copy of the final framework and were interested in applying it in their respective contexts straightaway. After phase 1 interviews, some additional details were added to the framework based on the feedback received from educators, ed-tech experts and AI practitioners.

*Table 1: Three groups of people interviewed for this research*

| Group | Description | Number of Candidates |
|---|---|---|

| Educators | Teachers, Principals and other Leaders in Schools | 9 |
|---|---|---|
| **Ed-tech Experts** | People leading the digital strategy initiatives and ed-tech implementations in schools | 5 |
| **AI Practitioners** | AI Practitioners | 4 |

*3.2.2 Data Collection Tool: Interviews*

The interviews with candidates were semi-structured and varied slightly between different groups. For example, with educators a high-level purpose of the framework was shared with minor details on how the framework evaluated AI tools throughout their development pipeline. They were shown what the framework would inform them about a particular AI product and then inquired if they will find such information useful. They were also asked if they have inquired about this information in the past and would they use a framework like this to audit the AI-powered ed-tech products before deploying them in their institutions.

With ed-tech experts, some details of the framework were shared, and they were inquired about the usefulness of this framework as an auditing tool to evaluate AI-powered ed-tech products. They were also inquired about their opinions on the impact of AI in education, its potential, its impact and any harms it could cause to learners and educators. For the ed-tech experts who were already using AI-powered ed-tech in their schools, they were also asked if they have ever had any conversations on AI ethics or transparency in AI with their ed-tech providers.

With AI practitioners, all the details of this framework were shared, and their opinions were also incorporated to improve the framework along with educators and ed-tech experts. They were inquired about the ethical considerations in place for their current projects and if they were already using any systematic processes for auditing their AI powered ed-tech tools. They were also asked if they have received any demands from schools for ensuring ethical AI development and how they address these requirements.

*3.2.3 Data Analysis*

After phase 1 interviews, explorative thematic analysis was conducted manually, and preliminary codes were assigned to the collected data from different interviews. Then patterns/themes were identified and reviewed across the assigned codes from interviews from each specific group (educators, ed-tech experts and AI practitioners) and from all the groups combined to take account of the unique requirements of each group. The Transparency Index framework was improved based on the findings from phase 1 interviews. Then phase 2 interviews were conducted followed by the deductive thematic analysis to confirm the findings from phase 1. The findings from phase 1 and phase 2 interviews were incorporated in the final version of the framework

## 4. RESULTS

In this section we present the results of this research in the form of the final version of the Transparency Index framework that was built based on the literature review, development of an AI-powered ed-tech tool and after the interviews with educators, ed-tech experts and AI practitioners.

### 4.1 Framework

The framework adopts a continuous approach where transparency is not seen as an instantaneous decision or is not dependent on the usage of a particular set of tools only. It is adopted as a continuous process, integrated into the design methodology throughout an AI tool's planning, development, deployment and usage scenarios. There are several factors that influence the type of transparency that should be or can be induced in an AI system. Some of these factors are shown in table 2 below:

We understand that the AI development process can be extremely complex and every AI-powered ed-tech product is unique with its own development and usage dynamics. The process through which transparency is ensured in an AI system or the extent to which transparency is needed for an AI system is determined by a number of factors that should be taken into account before starting the data collection process for an AI system. Some of these factors are as follows:

1. What approach will you adopt in deploying the AI system to production:
    1.1. Human in the Loop: Final decision-making authority is kept with the human, the AI system's role is to enable more informed decision making
    1.2. Human on the Loop: Human plays a supervisory role to evaluate the decisions made by an AI system before they are implemented in real-world
    1.3. Human out of the Loop: AI system makes the decision with no human involvement

2. What kind of impact will this AI system have on its users:

    2.1. Direct Impact: where a user's personal life is affected by a decision and in some cases the user has no choice other than compliance, like recruiting a candidate, giving a loan or deciding on recidivism
    2.2. Indirect Impact: where a user's day to day life is not affected by an AI system's decision or user has the choice to decide against the AI's decision, like spam filtering in emails, or recommendations on an e-commerce store

3. What is the tech-savviness of the individuals who would be (directly or indirectly) impacted by this AI system:
    3.1. Tier 1: Researchers or Practitioners: they have a thorough understanding of the techniques that are needed for the development of an AI system
    3.2. Tier 2: Software engineers and tech enthusiasts: they have some understanding of the techniques that are needed for the development of an AI system
    3.3. Tier 3: General public: they do not have any understanding of the techniques that are needed for the development of an AI system

Some of the requirements for transparency in each stage of the AI development process are as follows:

1. Data Transparency
    1.1. Data Collection
        1.1.1. How was data collected
        1.1.2. What were its sources
        1.1.3. Was Assumption Testing carried out: What assumptions were made regarding the data collection
            1.1.3.1. How many of these assumptions were tested and verified?
        1.1.4. Was consent taken from all individuals
        1.1.5. What data on sensitive variables are collected
        1.1.6. How is your data labelled:
            1.1.6.1. By ground-truths
            1.1.6.2. By human labels
        1.1.7. From 1 to 10, how do you rate the involvement of domain experts in data collection
    1.2. Data Processing
        1.2.1. How is data stored and ensured that it is secure
        1.2.2. How was data normalized
        1.2.3. What techniques and tools were used to process the data
            1.2.3.1. Were the strengths and weaknesses of these techniques explored
        1.2.4. How was the sensitive variables data processed
    1.3. Data Analysis
        1.3.1. What techniques and tools were used to analyze the data
        1.3.2. Was Exploratory data analysis done
            1.3.2.1. Was Exploratory data analysis shared and confirmed with domain experts
        1.3.3. Was statistical data analysis done
            1.3.3.1. Was statistical data analysis shared and confirmed with domain experts
        1.3.4. Was correlations between different features identified and confirmed by domain experts to evaluate any assumptions made
        1.3.5. What types of biases were identified in the data
            1.3.5.1. Historical Bias: This bias exists in the society and is reflected in the data even if there are no errors in the data collection and processing stages
            1.3.5.2. Representation Bias: This bias occurs when sample data used to build the AI is not truly representative of the real-world
            1.3.5.3. Measurement Bias: This bias occurs while choosing, collecting and computing features in the data to measure a certain outcome
            1.3.5.4. Aggregation Bias: This bias occurs when one size fits all AI approach is used for all the groups in the data
            1.3.5.5. Evaluation Bias: This occurs when the test data of an algorithm does not represent the target population
            1.3.5.6. Deployment Bias: This occurs when there is a mismatch between the problem that the AI tool is built to solve and what it is actually used for in the real- world
        1.3.6. What steps were taken to mitigate the above biases in the data
        1.3.7. Are domain experts informed of the measures taken to mitigate these biases?
        1.3.8. Are domain experts fully briefed on the potential impact of each type of bias on the AI system's predictions?
        1.3.9. Was Datasheet prepared for the data processing stage

2. Algorithmic Transparency
    2.1. Model Selection
        2.1.1. Which ML model was used for predictions
        2.1.2. Why was this particular model chosen?

   2.1.3. Was a Models Evaluation Report prepared: Was there any experimentation done with different machine learning models?
   2.1.4. What are some common strengths and weaknesses of this model
   2.1.5. In choosing the model were Transparency capabilities of the model taken into consideration
   2.1.6. Is the model doing regression or classification (classification is considered comparatively less risky in terms of bias as we're predicting a range rather than an exact value)
   2.1.7. Is the model using any Explainable Artificial Intelligence (XAI) tools or providing explanations of the predictions
     2.1.7.1. If yes, which XAI tools are being used
     2.1.7.2. What are the strengths and weaknesses of these tools
     2.1.7.3. Was human in the loop trained regarding the limitations of these explanations or
   2.1.8. Were any measures taken to address these limitations in autonomous AI systems
  2.2. Model Training
   2.2.1. Which tools (like Python libraries) were used for training the models
   2.2.2. What hyperparameters were used for training the models
     2.2.2.1. Were these hyperparameters optimized?
       2.2.2.1.1. If yes, what techniques were used for hyperparameter optimization
   2.2.3. What was the percentage of training and test set?
   2.2.4. Was the distribution of features in training and test set similar
  2.3. Model Verification
   2.3.1. How was the machine learning model audited. For example, what were the results of using counterfactuals etc
   2.3.2. Have you tested your model on a subset of sensitive variables data?
   2.3.3. Have you prepared a disclaimer document highlighting the exact contexts in which your model can be used
   2.3.4. Was a Model Card prepared for your models deployed in real-world?

3. Implementation Transparency
  3.1. AI Deployment
   3.1.1. Have you tested an MVP of this AI system with potential real-world users
     3.1.1.1. Were the domain experts satisfied with the tool's performance
     3.1.1.2. Will you share the details of this MVP testing with prospective clients
   3.1.2. Have you improved the tool based on that?
   3.1.3. Is there some form of visual signaling to indicate that a particular aspect of this AI system is work-in-progress, or are not perfect or have certain biases against these particular groups
  3.2. AI Monitoring
   3.2.1. How will you be monitoring this AI system in the production
   3.2.2. What security and privacy measures were taken when deploying the AI system
   3.2.3. Have you prepared a Models Validation Report to document the tool's performance in real-world with focus groups or the first few users
     3.2.3.1. Were the results up to expectation?
       3.2.3.1.1. If not, what changes were made in the AI system
       3.2.3.1.2. Were steps 3.2.3. onwards repeated unless the AI system reached expected results
  3.3. AI Improvements
   3.3.1. How often are you planning on pushing the improved model to production
   3.3.2. Have you identified the lower limit below which the AI system needs attention or human intervention
   3.3.3. Have you identified the lower limit below which the AI system should stop working
   3.3.4. Have you completed registration or acquired endorsements (like completing conformity assessments) from regulators or other third parties, like registration on public EU database for high-risk AI systems
   3.3.5. Have you prepared the Factsheet for your AI tool

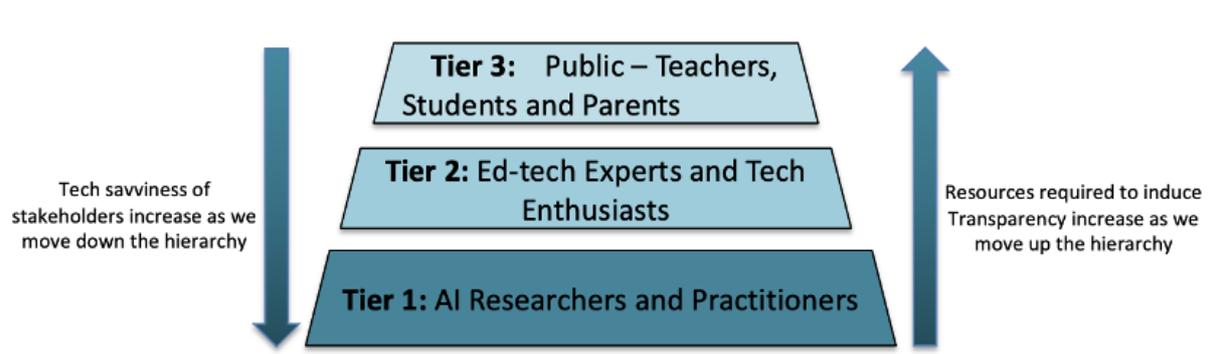

*Figure 4: Three Tiers of Transparency for AI in Education*

Figure 4 above maps the different types of stakeholders in terms of their technical background to the resources required to make an AI system transparent for that particular group. It shows how the requirements for transparency in AI products vary with the background of stakeholders. To make an AI system transparent for stakeholders with no technical background, like educators, AI companies need to invest significant resources and time.

*Table 2: Three Tiers of Transparency*

| Tiers | Description |
| --- | --- |
| Tier 1 | An AI system is considered Transparent for AI researchers and practitioners if they know:<br>· The Machine Learning model used in the AI system<br>· Optimization techniques used, like cross-validation or bootstrapping<br>· Hyperparameter optimization techniques used<br>· Hyperparameter values used<br>· Any particular opensource software, libraries or packages used<br>· Detailed documentation of the data processing and engineering<br>· The trained parameters values |
| Tier 2 | An AI system is considered Transparent for ed-tech experts and enthusiasts if it has all the above details plus:<br>· Explanations implemented in the AI system show which factors played the most important role for a particular prediction<br>· The AI system has a human in the loop who understands the explanations<br>· Technical understanding of human-in-the-loop is known<br>· Accuracy metrics of the ML model used in the AI system are shared<br>· Information about the distribution of sensitive variables like gender, race, religion in the data is shared<br>· Details of different third-party tools used in the AI system are shared, like Google Translate for translations or IBM Watson for conversations<br>· Detailed documentation of the data used like datasheets and models used like model cards (the assumptions made in the AI development process and disclaimers on the contexts in which this AI system cannot be used) |
| Tier 3 | An AI system is considered Transparent for educators and parents if it has all the above details plus:<br>· The AI system has been thoroughly tested with sample users and their findings have been incorporated into the product development and training<br>· AI Explanations are implemented in the AI system in the form of sentences that are easily understandable by the general public with minimal technical jargon<br>· Human in the loop is fully responsible for final decision-making<br>· Human in the loop understands the weaknesses of the data used for training ML models<br>· Human in the loop can explain the workings of an AI system to users<br>· Human in the loop has a thorough understanding of when to rely on an AI system and when to avoid it<br>· Human in the loop receives a training session on:<br>  o How to use the AI system<br>  o What are its weaknesses<br>  o Where can it go wrong<br>· User Interface of the AI system takes account of the distribution of sensitive variables in the data. Predictions for under-represented groups illustrate more information about what the training data lacks about such groups that might skew the results for a particular individual<br>· Weaknesses of different third-party tools used in the AI system are shared, like Oromo spoken in Ethiopia can't be translated correctly through google translate<br>· Carbon footprint of the energy used in training the models for the AI system is shared |

### 4.2 Emerging Themes of Feedback in Stakeholder Interviews

There were some themes that appeared across all the stakeholders from different groups, irrespective of their background. In contrast, there were some local themes that appeared consistently across all stakeholders from a particular group, with a specific background. For example, some themes emerged from educators with a minimal technical background but not from AI practitioners, or some themes appeared in conversations with ed-tech experts but not with AI practitioners.

a. **The value of the proposed framework**

Across all the groups, stakeholders thought that the framework was useful in enhancing their understanding of AI products and where these products can go wrong. Educators stated that the framework could help them get a better understanding of AI-powered ed-tech products and the contexts in which they work best. Some educators who were currently evaluating ed-tech tools were very interested in using the framework as an auditing tool to get a better understanding of these products.

Ed-tech experts also viewed this framework as an auditing tool for evaluating the ed-tech products before they are deployed in schools. They relied on GDPR regulations to ensure data privacy but were not aware of any tools or frameworks for AI ethics that are applicable to education.

AI practitioners perceived the framework as a documentation tool to record all the details of the machine learning development pipelines of the AI products they build. AI ethics was a concern for them but their applications of AI ethics on products were mostly limited to explanations. From the interviews, it seemed this was the first time they were looking at transparency at a detailed level on every stage of the machine learning development pipeline.

### b. Transparency of AI products is a relatively new phenomenon for stakeholders

For 17 out of the 18 stakeholders interviewed for this research, the conversation on transparency regarding ethical considerations in AI products was a relatively new phenomenon. Seven educators were using some form of ed-tech and AI products in schools, but ethical AI was a relatively new conversation for them. It seemed they were not aware of the adverse consequences of AI going wrong or not working as expected.

All eighteen interviewees, including most of the AI practitioners who build AI products did not perceive transparency in machine learning development pipelines the way it was being addressed by the proposed framework in this research. Ed-tech experts seemed to rely on government regulations for protection against AI mishaps. The issues with this approach were that firstly, there are no clear government regulations regarding the bias in AI systems or their malfunctioning, and secondly the government intervention may occur after the damage has been done.

### c. Focus on Transparency and Ethics in AI products

The concept of making their AI implementations transparent throughout the machine learning development pipeline and sharing it with end-users was new to AI practitioners.

They made an interesting point regarding transparency in AI stating that they do not put in a lot of effort in making their AI products transparent because their clients (education institutions like schools) do not ask for it. This theme also emerged in our interviews with educators and ed-tech experts in schools: they were mostly not aware of the importance of transparency in the AI products they use in schools. Although one practitioner confirmed that in the past they have received questions from teachers when teachers noticed something strange, like why the tool is making this particular decision for this student.

All four AI practitioners that were interviewed claimed that they try to make their machine learning models explainable. They were not concerned about the understandability of those explanations by users. They focus on post-hoc explainability through tools like Lime (Ribeiro et al, 2016). But they have never received particular requests from educational institutions on adding explainability to their AI products or sharing the details of their development process.

Ed-tech experts identified the importance of ed-tech and AI in ed-tech to enhance the learning outcomes for students. In their opinion, the claim that AI powered ed-tech products reduce teacher workload is not backed by evidence. They claimed that AI in ed-tech is over-rated and not as impactful as some companies claim.

When asked about their concerns regarding AI going wrong and negatively impacting learning outcomes based on false information or wrong predictions, two contrasting themes emerged:
- Ed-tech experts relied on regulations like GDPR to take care of any such mishaps. Ed-tech companies operating within the EU need to make their products compliant with GDPR. They thought that GDPR would also take care of any ethical, accountability and transparency issues within AI along with data privacy and storage concerns
- Another theme that emerged in the discussions with ed-tech experts focused on AI hype in ed-tech products. They thought that ed-tech companies claiming to use AI in their products exaggerated the benefits of AI. According to them, any major breakthroughs in AI would require huge amounts of investment. These ed-tech experts were also aware of the importance of ethical AI in education. They claimed that the unintended consequences of AI in education are not as well documented as in other sectors. This poses more danger as most educators, and sometimes even ed-tech companies are not very well aware of where AI can go wrong in the context of education and the impact this can have on learners. It is important to note that the ed-tech experts with these beliefs thought that there might have been mishaps in AI in education, but they are not very well documented.

For educators, transparency in AI was a new phenomenon. They were excited about trying new ed-tech and AI products in their schools and evaluating their impact on learning outcomes, but mostly did not seem concerned about AI's negative consequences. Though most of them were not tech savvy, they understood the purpose of this research's framework and how it could be useful in their contexts.

For all nine educators interviewed, this was the first time they were having conversations on ethical AI and what kind of documentation or precautionary measures to expect from companies applying AI in education. It seemed they never had these conversations with their ed-tech providers earlier. A theme that was also confirmed by an AI practitioner leading a data science team at one of the biggest ed-tech companies in the world.

### d. Feedback and Recommendations to Improve the Framework

AI practitioners were the most important group in terms of providing feedback to improve the framework. For example, they recommended adding a brief summary of each type of bias that can exist in a machine learning development pipeline. They also advised adding another clause regarding open-sourcing or publicly sharing the development code that was used to build the product. This would enable any watchdog or AI auditing group to replicate the results of that AI product and identify gaps.

AI practitioners showed the most interest in the details of the framework, especially the data processing stage. AI practitioners requested to view a copy of this framework and showed interest in using this research in their projects when it is published.

Most educators wanted to have follow-up conversations on the framework. They wanted to discuss it with their colleagues and incorporate some of the questions in teacher training sessions to enhance their understanding of AI products before they are used in classrooms. From the conversations, it seemed that the framework helped educators in identifying a gap in their current auditing process for AI-powered ed-tech products before they are deployed in schools.

Some educators also requested to view a copy of the framework to give feedback. One educator pointed out that the Transparency for Tier 3 users (general public: teachers, students and parents) 'is too esoteric and more linked to their perceived outcomes'. Based on their feedback, a separate version of the framework was prepared specifically for schools. This version of the framework will have definitions of all the technical terms like different types of bias.

Another educator requested that the language used in the framework needs to be 'explicated/simplified for ease of access'.

AI practitioners were also the most active group in terms of giving recommendations on how the framework could be improved. Most of their suggestions were accepted and the final version of the framework has been presented above.

## 5. DISCUSSION

For AI systems deployed in educational contexts, it's very important to thoroughly document the data they are trained on because this training data plays a significant role in the kind of contexts that an AI system would work expectedly. The Transparency Index framework provides different aspects of the data processing stage that AI practitioners need to document. For ML modelling, the details of the models used for decision-making can be documented and shared. To take this one step further, the ed-tech companies developing AI-powered products can also choose to publicly share the code that their AI practitioners and data scientists write. This can enable reproducible research and contribute to the tech community working on AIED.

Some might argue that steps like making the code of AI implementations public through GitHub or other tools is not very helpful for the general public or Tier 3 users mentioned in the framework. But, such steps help the tier 3 users indirectly. This has both a push and a pull factor for AI practitioners as they know their work (code) will be visible to the public in future which reinstates the need to work towards the public good (Elster; 1998; Chambers, 2004; Chambers 2005; Naurin, 2007). It also means that practitioners know they can be held accountable for their work.

One of the themes that emerged across all three stakeholders of AI in education, including educators, ed-tech experts and AI practitioners focused on how all the groups found transparency through the Transparency Index framework useful. Educators as the users of AI-powered ed-tech products and AI practitioners as developers and providers of AI systems in education believed transparency helped them in understanding their products better. But some researchers have also argued against complete transparency like Zarski (2016), Lepri et al (2017), De Laat (2018) and Carabantes (2019). Complete transparency like making the code of an AI tool public can hinder innovation as companies will be sharing the secret sauce that makes AI work in certain contexts and provides them with a competitive edge over others.

Another problem with complete transparency as shown in the Transparency Index framework is that it can potentially lead to information overload for stakeholders (Eppler and Mengis, 2004) or transparency paradox (Richards and King, 2013). Sharing everything with the stakeholders can potentially confuse them and make it more difficult for them to find the relevant information (Stohl et al, 2016). Some researchers like Heald (2006) have used the term 'transparency illusion' to illustrate this phenomenon. AI-powered ed-tech companies for teachers face this risk as teachers in a classroom setting can be easily overloaded with too much information on their dashboards (Bull et al, 2013; Greller and Drechsler, 2012). In such scenarios it can be argued that the ed-tech company developing an AI product can implement the different components of the Transparency Index framework for all three stages of the AI tool development process but can avoid sharing all this information with end-users or third parties.

Luckin *et al* (2006) have illustrated the importance of human-centered design in developing educational systems that are fit for use. They highlight the importance of iterative improvements in building such educational systems. Considering the risks involved, this particularly holds true for AI systems in education. It is very important to thoroughly test these systems with sample users before deploying them at scale. Different components of the 'Implementation Transparency' section in the Transparency Index framework tend to address these concerns.

It can be argued that many times users may not even know what information they need. What is useful for them, what kind of impact lack of transparency can have on them or what is too much transparency for them that leads to cognitive overload. This is especially the case for tier 3 users who are not tech experts and do not know exactly what kind of information from the entire AI tool's development pipeline will be useful for them. This is also shown in the findings of this study that educators who are also tier 3 users (according to the framework) have mostly never had conversations on ethical AI and/or transparency in AI before. AI practitioners also confirmed this by saying that they have never received requests for transparency or concerns about ethical aspects of AI in product development from their clients (educators).

A counterpoint to the above argument is that even if sharing the development details of an AI tool leads to cognitive overload, this does not mean that ed-tech companies should stop making such information public at all. End-users of an AI tool do not necessarily need to know or understand every detail of an AI implementation, but this belief in AI practitioners that they need to share every decision and assumption made during the tool's development can act as a strong precautionary measure for them to double-check these decisions, leading to more robust development processes. These checks and balances can also prevent mistakes that lead to controversial results and can be harmful to the ed-tech company's image.

The criticisms on transparency are mostly directed at the information that is shared with end-users. If the focus is on the question of 'transparency for whom' and the transparency measures to be taken by ed-tech companies when developing AI are treated separately from the information that they need to share with various stakeholders of education, then it can be noticed that the above critique is mostly directed at the information shared with end-users, not the measures to be taken by ed-tech companies. For example, the autopilots working in cars are powered by state-of-the-art image recognition algorithms trained on vast amounts of data (Hirz and Walzel; 2018). When drivers are using the autopilots, they do not necessarily want to know the details of how AI is making every decision or identifying different road signals etc. They need to know when not to trust the AI system, for example during heavy rainfall etc. But, if the company developing this software does not feel the need to share the details of their AI system with end-users, it does not imply that they should ignore the transparency considerations while developing that AI system. Transparency may lead to more robust and well-documented AI systems. Therefore, the documentation of the decisions taken, and assumptions made during the AI tool development process can be valuable for the company itself (Madaio et al, 2020).

Cognitive overload in the context of transparency is caused by sharing too much information with the users of an AI system (Kirsh, 2000) like teachers, headteachers and learners. There are a number of ways in which cognitive overload for such users can be avoided without compromising on the principles of transparency. For example, despite the documentation of the entire AI tool's development pipeline only relevant information can be shared with the stakeholders. If this information is too much, it can be shared over a period of time or made available to stakeholders and left at their discretion to access it as and when needed. For example, Cukurova *et al* (2019) have presented a framework for evidence-informed educational technology where ed-tech companies work closely with researchers and educators to ensure the efficacy of the products they build. Evidence may not always be fed into stakeholders, but should be available, when/if a practitioner requests to have access to it. Similarly, for safety in AI systems, a participatory design methodology where ed-tech companies closely work with the prospective users of their AI offering to understand their needs is necessary (Luckin et al, 2011).

6. **LIMITATIONS AND FUTURE WORK**

In future, the Transparency Index framework for education proposed in this research can be further developed into a scoring system to evaluate AI-powered ed-tech products. This system could act as an indicator regarding the steps that an ed-tech company developing an AI-powered product has taken to ensure ethical AI development.

Future work on transparency in AI can also focus on how the Transparency Index framework proposed in this research can be adopted in other industries like healthcare, financial services or judiciary. It would be interesting to note the changes that the framework goes through across different sectors.

One of the limitations of the framework is that it was evaluated with different stakeholders of AI in education only within the United Kingdom. It is possible for the framework to not work as effectively in other locations like Asia and North America with different regulations on personal data collection, where adoption of AI and ed-tech in schools is not the same as in the UK and the curriculum and culture of schools vary significantly.

7. CONCLUSION

The Transparency Index framework proposed in this research integrates the popular frameworks of ethical AI like Datasheets, Model Cards and Factsheets into one coherent framework that addresses the whole AI product development timeline for Educational Technology interventions. We contextualize these tools in a single framework for their applicability in educational contexts and validate these modifications through interviews with various stakeholders of AI in education.

The Transparency Index is a comprehensive framework to evaluate, audit and analyze the effectiveness of AI systems in education. It can be utilized by different stakeholders of AI in education including educators, teachers, learners, ed-tech experts, executive leaders and AI practitioners developing ed-tech products. Educators can utilize this framework to evaluate the AI-powered ed-tech being used in their schools, AI practitioners can use this as a checklist to document the robustness of their AI development processes and ed-tech experts can use the Transparency Index framework as an auditing tool before recommending an AI-powered ed-tech product.

Recently, there has been significant research work on developing checklists and frameworks for ethical AI. This research takes it forward by proposing a robust framework for transparency in AI systems applied in education. It shows how AI practitioners and ed-tech companies developing AI-powered products can make sense of the measures they take to ensure ethical AI for different tiers of stakeholders. It also highlights the importance of Transparency for companies to develop robust and ethical AI development pipelines and for stakeholders to get a better understanding of how the AI systems that impact them actually work.

8. REFERENCES

1. AlgorithmWatch. (2019). Automating society: Taking stock of automated decision-making in the EU. Retrieved June 29, 2020, from https://algorithmwatch.org/wp-content/uploads/2019/01/Automating_Society_Report_2019.pdf.
2. Andrew D Selbst, Danah Boyd, Sorelle A Friedler, Suresh Venkatasubramanian, and Janet Vertesi. Fairness and abstraction in sociotechnical systems. In Proceedings of the Conference on Fairness, Accountability, and Transparency, pages 59–68. ACM, 2019.
3. Arnold, M., Bellamy, R.K., Hind, M., Houde, S., Mehta, S., Mojsilović, A., Nair, R., Ramamurthy, K.N., Olteanu, A., Piorkowski, D. and Reimer, D., 2019. FactSheets: Increasing trust in AI services through supplier's declarations of conformity. *IBM Journal of Research and Development*, *63*(4/5), pp.6-1.
4. Brill J (2015) Scalable approaches to transparency and accountability in decisionmaking algorithms: remarks at the NYU conference on algorithms and accountability. Federal Trade Commission, 28 February. Available at: https://www.ftc.gov/system/files/documents/public_statements/629681/150228nyualgorithms.pdf
5. Bughin, Jacques, Jeongmin Seong, James Manyika, Michael Chui, and Raoul Joshi. "Notes from the AI frontier: Modeling the impact of AI on the world economy." *McKinsey Global Institute* (2018).
6. Bull, S., Kickmeier-Rust, M., Vatrapu, R.K., Johnson, M.D., Hammermueller, K., Byrne, W., Hernandez-Munoz, L., Giorgini, F. and Meissl-Egghart, G., 2013, September. Learning, learning analytics, activity visualisation and open learner model: Confusing?. In *European Conference on Technology Enhanced Learning* (pp. 532-535). Springer, Berlin, Heidelberg.
7. Carabantes M (2019) Black-box artificial intelligence: an epistemological and critical analysis. AI Soc. https://doi.org/10.1007/s0014 6-019-00888-w
8. Chambers S (2004) Behind closed doors: publicity, secrecy, and the quality of deliberation. J Polit Philos 12(4):389–410
9. Chambers S (2005) Measuring publicity's effect: reconciling empirical research and normative theory. Acta Polit 40(2):255–266
10. Crawford, K., Whittaker, M., Elish, M.C., Barocas, S., Plasek, A. and Ferryman, K., 2016. The AI Now Report. *The Social and Economic Implications of Artificial Intelligence Technologies in the Near-Term.*
11. Cukurova, M., Luckin, R. and Clark-Wilson, A. (2018). Creating the golden triangle of evidence-informed education technology with EDUCATE. *British Journal of Educational Technology*, 50(2), pp.490–504.
12. Cukurova, M., Luckin, R. and Kent, C., 2020. Impact of an artificial intelligence research frame on the perceived credibility of educational research evidence. *International Journal of Artificial Intelligence in Education*, *30*(2), pp.205-235.

13. Cukurova, M., Luckin, R., & Clark-Wilson, A. (2019). Creating the golden triangle of evidence-informed education technology with EDUCATE. *British Journal of Educational Technology*, *50*(2), 490-504.
14. Dameski, A., 2018, August. A comprehensive ethical framework for AI entities: Foundations. In *International Conference on Artificial General Intelligence* (pp. 42-51). Springer, Cham.
15. de Fine Licht, K. and de Fine Licht, J., 2020. Artificial intelligence, transparency, and public decision-making. *AI & SOCIETY*, pp.1-10.
16. De Laat PB (2018) Algorithmic decision-making based on machine learning from Big Data: Can transparency restore accountability? Philos Technol 31(4):525–541
17. DeepMind Safety Research (2018). Building safe artificial intelligence: specification, robustness, and assurance. [online] Medium. Available at: https://medium.com/@deepmindsafetyresearch/building-safe-artificial-intelligence-52f5f75058f1.
18. Diakopoulos N (2016) Accountability in algorithmic decision making. Communications of the ACM 59(2): 56–62.
19. Dubber, M.D., Pasquale, F. and Das, S. eds., 2020. *The Oxford Handbook of Ethics of AI*. Oxford University Press, USA.
20. Elster J (1998) Deliberation and constitution making. In: Elster J (ed) Deliberative Democracy. Cambridge University Press, Cambridge
21. Eppler MJ, Mengis J (2004) The concept of overload: a review of literature from organization science, accounting, marketing, MIS, and related disciplines. Inf Soc 20(5):325–344
22. European Commission. (2020). White paper on artificial intelligence: A European approach to excellence and trust. Retrieved August 19, 2020, from https://ec.europa.eu/info/sites/info/files/commission-white-paper-artificial-intelligence-feb2020_en.pdf.
23. Felzmann, H., Fosch Villaronga, E., Lutz, C., & Tamò-Larrieux, A. (2019b). Robots and transparency: The multiple dimensions of transparency in the context of robot technologies. *IEEE Robotics and Automation Magazine*, *26*(2), 71–78.
24. Felzmann, H., Fosch-Villaronga, E., Lutz, C. and Tamò-Larrieux, A., 2020. Towards transparency by design for artificial intelligence. *Science and Engineering Ethics*, pp.1-29.
25. Floridi, L., Cowls, J., Beltrametti, M., Chatila, R., Chazerand, P., Dignum, V., Luetge, C., Madelin, R., Pagallo, U., Rossi, F., Schafer, B., Valcke, P. and Vayena, E. (2018). AI4People—An Ethical Framework for a Good AI Society: Opportunities, Risks, Principles, and Recommendations. *Minds and Machines*, 28(4), pp.689–707.
26. Gebru, T., Morgenstern, J., Vecchione, B., Vaughan, J.W., Wallach, H., Daumé III, H. and Crawford, K., 2018. Datasheets for datasets. *arXiv preprint arXiv:1803.09010*.
27. Greller, W. and Drachsler, H., 2012. Translating learning into numbers: A generic framework for learning analytics. *Journal of Educational Technology & Society*, *15*(3), pp.42-57.
28. Heald D (2006) Varieties of transparency. Proceedings of the British Academy 135: 25–43.
29. Hirz, M. and Walzel, B., 2018. Sensor and object recognition technologies for self-driving cars. *Computer-aided design and applications*, *15*(4), pp.501-508.
30. Hollanek, T. (2020). AI transparency: a matter of reconciling design with critique. *AI & SOCIETY*.
31. Holmes, W., Bektik, D., Di Gennaro, M., Woolf, B. P., & Luckin, R. (2019a). Ethics in AIED: Who cares? In S. Isotani, E. Millán, A. Ogan, P. Hastings, & R. Luckin (Eds.), Artificial Intelligence in Education (Vol. 11625, pp. 424–425). Cham, Switzerland: Springer International Publishing AG. https://doi.org/10.1007/978-3-030-23204-7.
32. Holmes, W., Bialik, M., & Fadel, C. (2019b). Artificial intelligence in education. Promises and Implications for Teaching and Learning. Center for Curriculum Redesign.
33. Holmes, W., Porayska-Pomsta, K., Holstein, K., Sutherland, E., Baker, T., Shum, S.B., Santos, O.C., Rodrigo, M.T., Cukurova, M., Bittencourt, I.I. and Koedinger, K.R. (2021). Ethics of AI in Education: Towards a Community-Wide Framework. *International Journal of Artificial Intelligence in Education*.
34. Holstein, K., Wortman Vaughan, J., Daumé III, H., Dudik, M. and Wallach, H., 2019, May. Improving fairness in machine learning systems: What do industry practitioners need?. In *Proceedings of the 2019 CHI conference on human factors in computing systems* (pp. 1-16).
35. Hood C (2010) Accountability and transparency: siamese twins, matching parts, awkward couple? West Eur Polit 33(5):989–1009
36. IBM. (2018). *Principles for trust and transparency*. Retrieved November 11, 2020, from https://www.ibm.com/blogs/policy/wp-content/uploads/2018/05/IBM_Principles_OnePage.pdf. Accessed.
37. ICDPPC (International Conference of Data Protection and Privacy Commissioners). (2018). *Declaration on ethics and data protection in artificial intelligence*. Retrieved November 11, 2020, from https://icdppc.org/wp-content/uploads/2018/10/20180922_ICDPPC-40th_AI-Declaration_ADOPTED.pdf.
38. IEEE. (2019). *Ethically aligned design* (version 2). Retrieved November 11, 2020, from https://standards.ieee.org/content/dam/ieee-standards/standards/web/documents/other/ead_v2.pdf.
39. Jarrahi, M.H., 2018. Artificial intelligence and the future of work: Human-AI symbiosis in organizational decision making. *Business Horizons*, *61*(4), pp.577-586.
40. Jobin, A., Ienca, M. and Vayena, E., 2019. The global landscape of AI ethics guidelines. *Nature Machine Intelligence*, *1*(9), pp.389-399.

41. K. Hänsel, N. Wilde, H. Haddadi, A. Alomainy, 2015, Challenges with current wearable technology in monitoring health data and providing positive behavioural support, in: Proceedings of the 5th EAI International Conference on Wireless Mobile Communication and Healthcare, pp. 158–161.
42. Kaushal, A., Altman, R. and Langlotz, C., 2020. Health care AI systems are biased. *Scientific American*.
43. Kazim, E. and Koshiyama, A., 2020. A high-level overview of AI ethics. *Available at SSRN*.
44. Kent, C., Chaudhry, M.A., Cukurova, M., Bashir, I., Pickard, H., Jenkins, C., du Boulay, B., Moeini, A. and Luckin, R. (2021). Machine Learning Models and Their Development Process as Learning Affordances for Humans. *Lecture Notes in Computer Science*, pp.228–240.
45. Kirsh, D., 2000. A few thoughts on cognitive overload. *Intellectica, 1*(30).
46. Larsson, S. and Heintz, F., 2020. Transparency in artificial intelligence. *Internet Policy Review, 9*(2).
47. Leikas, J., Koivisto, R. and Gotcheva, N., 2019. Ethical framework for designing autonomous intelligent systems. *Journal of Open Innovation: Technology, Market, and Complexity, 5*(1), p.18.
48. Lepri B, Oliver N, Letouzé E, Pentland A, Vinck P (2017) Fair, transparent, and accountable algorithmic decision-making processes. Philos Technol 2017:1–17
49. Linardatos, P., Papastefanopoulos, V. and Kotsiantis, S., 2021. Explainable ai: A review of machine learning interpretability methods. *Entropy, 23*(1), p.18.
50. Luckin, R., 2018. *Machine Learning and Human Intelligence: The future of education for the 21st century*. UCL IOE Press. UCL Institute of Education, University of London, 20 Bedford Way, London WC1H 0AL.
51. Luckin, R., 2021. What inspired my thinking to create UCL EDUCATE?. *Research for All*.
52. Luckin, R., Clark, W., Garnett, F., Whitworth, A., Akass, J., Cook, J., Day, P., Ecclesfield, N., Hamilton, T. and Robertson, J., 2011. Learner-generated contexts: A framework to support the effective use of technology for learning. In *Web 2.0-based e-learning: applying social informatics for tertiary teaching* (pp. 70-84). IGI Global.
53. Luckin, R., Underwood, J., du Boulay, B., Holmberg, J., Kerawalla, L., O'Connor, J., Smith, H. and Tunley, H. (2006). Designing Educational Systems Fit for Use: A Case Study in the Application of Human Centred Design for AIED
54. Madaio, M.A., Stark, L., Wortman Vaughan, J. and Wallach, H., 2020, April. Co-designing checklists to understand organizational challenges and opportunities around fairness in ai. In *Proceedings of the 2020 CHI Conference on Human Factors in Computing Systems* (pp. 1-14).
55. Matthias, A. (2004). The responsibility gap: Ascribing responsibility for the actions of learning automata. *Ethics and Information Technology, 6*(3), 175–183.
56. Mike Ananny and Kate Crawford. Seeing without knowing: Limitations of the transparency ideal and its application to algorithmic accountability. new media & society, 20(3):973–989, 2018.
57. Mitchell, M., Wu, S., Zaldivar, A., Barnes, P., Vasserman, L., Hutchinson, B., Spitzer, E., Raji, I.D. and Gebru, T., 2019, January. Model cards for model reporting. In *Proceedings of the conference on fairness, accountability, and transparency* (pp. 220-229).
58. Moeini, A., 2020. *Theorising Evidence-Informed Learning Technology Enterprises: A Participatory Design-Based Research Approach* (Doctoral dissertation, UCL (University College London)).
59. Morley, J. and Floridi, L., 2020. An ethically mindful approach to AI for health care. *Available at SSRN 3830536*.
60. Naurin D (2007) Deliberation behind closed doors: transparency and lobbying in the European Union. ECPR Press, Colchester
61. Pasquale F (2015) The Black Box Society: The Secret Algorithms That Control Money and Information. Cambridge, MA: Harvard University Press.
62. Phillips-Wren, G., 2012. AI tools in decision making support systems: a review. *International Journal on Artificial Intelligence Tools, 21*(02), p.1240005.
63. Plale, B., 2019, September. Transparency by Design in eScience Research. In *2019 15th International Conference on eScience (eScience)* (pp. 428-431). IEEE.
64. Raji, I.D. and Buolamwini, J., 2019, January. Actionable auditing: Investigating the impact of publicly naming biased performance results of commercial ai products. In *Proceedings of the 2019 AAAI/ACM Conference on AI, Ethics, and Society* (pp. 429-435).
65. Ribeiro, M.T., Singh, S. and Guestrin, C. (2016). *"Why Should I Trust You?": Explaining the Predictions of Any Classifier*. [online] arXiv.org. Available at: https://arxiv.org/abs/1602.04938.
66. Richards NM, King JH (2013) Three paradoxes of big data. Stan L Rev Online 66:41
67. Schiffer, S., 1991. Ceteris paribus laws. *Mind, 100*(1), pp.1-17.
68. Shen, J., Zhang, C.J., Jiang, B., Chen, J., Song, J., Liu, Z., He, Z., Wong, S.Y., Fang, P.H. and Ming, W.K., 2019. Artificial intelligence versus clinicians in disease diagnosis: systematic review. JMIR medical informatics, 7(3), p.e10010.
69. Smith, B. and Linden, G., 2017. Two decades of recommender systems at Amazon. com. *Ieee internet computing, 21*(3), pp.12-18.
70. Sterman, J.D., 2002. All models are wrong: reflections on becoming a systems scientist. *System Dynamics Review: The Journal of the System Dynamics Society, 18*(4), pp.501-531.
71. Stohl C, Stohl M and Leonardi PM (2016) Managing opacity: information visibility and the paradox of transparency in the digital age. International Journal of Communication Systems 10: 123–137.
72. The Institute for Ethical AI in Education The Ethical Framework for AI in Education. (n.d.). [online] . Available at: https://fb77c667c4d6e21c1e06.b-cdn.net/wp-content/uploads/2021/03/The-Institute-for-Ethical-AI-in-Education-The-Ethical-Framework-for-AI-in-Education.pdf

73. Turilli, M. and Floridi, L. (2009). The ethics of information transparency. *Ethics and Information Technology*, 11(2), pp.105–112.
74. Vaishya, R., Javaid, M., Khan, I.H. and Haleem, A., 2020. Artificial Intelligence (AI) applications for COVID-19 pandemic. *Diabetes & Metabolic Syndrome: Clinical Research & Reviews*, *14*(4), pp.337-339.
75. Weller, A. (2017). Challenges for transparency. arXiv preprint arXiv:1708.01870.
76. Wellner, G.P., 2020. When AI Is Gender-biased. *HUMANA. MENTE Journal of Philosophical Studies*, *13*(37), pp.127-150.
77. Wischmeyer, T., 2020. Artificial intelligence and transparency: opening the black box. In *Regulating artificial intelligence* (pp. 75-101). Springer, Cham.
78. Zarsky T (2016) The trouble with algorithmic decisions: an analytic road map to examine efficiency and fairness in automated and opaque decision making. Sci Technol Human Values 41(1):118–132
79. Završnik, A., 2020, March. Criminal justice, artificial intelligence systems, and human rights. In *ERA Forum* (Vol. 20, No. 4, pp. 567-583). Springer Berlin Heidelberg.